\documentclass[aps,prb,superscriptaddress,twocolumn,reprint,10pt,floatfix]{revtex4-1}
\usepackage{amsfonts}
\usepackage{amstext}
\usepackage{amssymb}
\usepackage{amsmath}
\usepackage{dsfont}
\usepackage[utf8]{inputenc}         
\usepackage{graphicx}
\usepackage{color,xcolor}
\usepackage{bm,stmaryrd}

\usepackage{hyperref}
\hypersetup{colorlinks=true,citecolor=blue,linkcolor=blue,filecolor=blue,urlcolor=blue}

\def\a{\alpha}

\def\G{\Gamma}
\def\D{\Delta}

\def\e{\varepsilon}

\bibpunct{[}{]}{,}{n}{}{}

\begin{document}

\title{Time-dependent resonant tunneling transport: Keldysh and Kadanoff-Baym nonequilibrium Green's functions in an analytically soluble problem}
\author{Mariana M. Odashima}
\affiliation{Instituto de F\'isica, Universidade Federal de Uberl\^andia, 38400-902 Uberl\^andia, MG, Brazil}
\email[Corresponding author: ]{mmodashima@ufu.br}
\author{Caio H. Lewenkopf}
\affiliation{Instituto de F\'isica, Universidade Federal Fluminense, 24210-346 Niter\'oi, RJ, Brazil}

\date{\today}

\begin{abstract}
Here we address two nonequilibrium Green's functions approaches for a resonant tunneling structure under a sudden switch of a bias. Our aim is to stress that the time-dependent Keldysh formulation of Jauho, Wingreen and Meir, and the partition-free scheme of Stefanucci and Almbladh are formally equivalent in the ubiquitous case of wide-band limit and noninteracting electrons, 
if leads and dot are  in equilibrium before the time-dependent perturbation.
We develop explicit closed formulas of the lesser Green's function and time-dependent current, reminding that the different 
integration limits preclude a face-to-face comparison of two approaches.
This study sheds light on both practices, which are of great interest to the mesoscopic transport community.
\end{abstract}
\pacs{PACS}

\maketitle

\section{Introduction}

Nonequilibrium Green's functions (NEGF) provide a solid basis for the theoretical understanding 
of the quantum electronic transport properties in a broad variety of systems \cite{HaugJauho}. 
The NEGF framework encompasses linear response as well as far from equilibrium conditions, 
yielding transient time-dependent  and/or steady-state currents, and two-time propagators as a 
function of the coupling and bias \cite{Jauho93, Jauho94, Stefanucci04, Schmidt, velickyPhysicaE, SpickaRev}. 
Nonetheless, despite the significant advances in treating out-of-equilibrium quantum many-body problems
\cite{Kamenev2011, Stefanucci-book}, 
further developments are needed in order to better account for relaxation mechanisms \cite{Eisert2015, Vasseur2016}, 
external perturbations \cite{Alexis,Croy2012, Eissing2016, Sanchez14}, and initial conditions \cite{Hall,Myo}. Recent efforts towards a nonequilibrium \textit{ab initio} theory 
and a unified contour picture have contributed to a growing interest on the subject 
\cite{Leeuwen13, Stefanucci-book}.

The Keldysh NEGF were introduced in the theory of transport through tunneling junctions by 
Caroli \textit{et al.} \cite{Caroli}, who developed a nonequilibrium perturbation theory assuming 
that the initial state consists of separate leads and a central region. In a remote past each partition 
is in equilibrium characterized by their own chemical potential. The full system is adiabatically 
connected by switching on the contact tunneling. 
The authors \cite{Caroli} alert that in this procedure the application of the bias happens before the different 
parts of the system are connected. The coupling corresponds to the time-dependent (switch on) 
perturbation.
Following these lines, an important contribution was achieved by Jauho, Wingreen and Meir (JWM) \cite{Jauho94}, 
who developed formal expressions for  resonant tunneling transport through an interacting region based 
on Green's functions on the Keldysh contour, for both steady-state and transient regimes
\cite{Jauho94}.

 More recently, alternative NEGF formulations have been explored to account for correlated initial conditions, relevant for short-time transients \cite{VelickyPRB}. The extended Keldysh contour, discussed at length for instance by Refs.\cite{Wagner, Danielewicz, KonstantinovPerel, KadanoffBaym} considers an initial state where the whole system is already in thermal equilibrium in the grand-canonical ensemble. An extended imaginary branch is added to the original contour, which starts at a time $t_0$, where Matsubara Green's functions describe the correlated initial state. 
 Stefanucci and Almbladh \cite{Stefanucci04} have obtained a closed formula for the lesser Green's function 
 of the central region in a lead-device-lead configuration using the extended Kadanoff-Baym contour, which 
 exhibits contributions due to the imaginary time convolutions that are apparently missing in the JWM approach. 
 Thus, for the out-of-equilibrium situation of a sudden switch-on of a bias, Stefanucci and Almbladh claimed to have developed an improved description of the transient currents over the Keldysh partitioned scheme of JWM \cite{Jauho93,Jauho94}. Ridley \textit{et al.} \cite{Ridley} have recently arrived at the same currents using both formalisms by taking the limit of $t_0\to-\infty$. However, in their understanding, only the steady-state would be reproduced, while transients would be missing from the partitioned scheme of JWM.

In this article we investigate these conflicting results by examining these two nonequilibrium Green's functions approaches for the double-barrier resonant tunneling system, the simplest prototype of a nanoelectronic device. At time $t_0$ we consider a sudden switch-on of a bias in one lead and develop the explicit expressions of the lesser Green's function and time-dependent current in the Keldysh and Kadanoff-Baym contours. For non-interacting electrons and in the wide flat band approximation, the problem is analytically soluble. In an effort to clarify the partitioning discussion in the literature, we address questions such as: Why do these two methodologies lead to the same results if the initial states are different? Why are the more general imaginary contour terms reproduced by the Keldysh approach? Can one stick to the state-of-the-art Keldysh NEGF for transients even beyond WBL and interacting electrons? 
Questions of this kind arise when extending the NEGF formalism to transients, therefore it is timely to identify very clearly the points of discrepancy or equivalence between the two present schemes.

This paper is structured as follows: in Sec. II we present the Jauho, Wingreen and Meir formulation of time-dependent resonant tunneling transport. In Sec.III the assumptions of the Keldysh approach are discussed and within this contour the time-dependent current is obtained. In Sec. IV we present the extended Keldysh contour and the results of lesser Green's function, followed by the conclusions.

\section{General formulation}

The model of time-dependent resonant tunneling transport we consider consists of a central region, such as a quantum dot, connected to the two metallic electrodes, described by the bilinear Hamiltonian \cite{Jauho94}

\begin{eqnarray} 
 H = \sum_{k,\alpha=L,R} \epsilon_{k\alpha}(t) c^{\dagger}_{k\alpha}c_{k\alpha} + \epsilon_0(t){d^\dagger_0}{d_0} + \nonumber \\
 + \sum_{k,\alpha=L,R} \left[ V_{k\alpha}(t) c^{\dagger}_{k\alpha}d_{0} + {\rm H.c.}\right] \,, \label{Ham}
\end{eqnarray}
where $c^{\dagger}_{k\alpha}(c_{k\alpha})$ creates (annihilates) an electron with momentum $k$ in the in the left ($\alpha=L$) or right ($\alpha=R$) lead, $d^{\dagger}_0(d_0)$ creates (annihilates) an electron at the resonance of energy $\e_0$ in the central region C and $V_{k\alpha}$ is the tunneling coupling parameter. For clarity we consider a single-level quantum dot (a noninteracting multilevel treatment is straightforward but algebraically involved). This approximation has the main advantage that the problem becomes analytically soluble, allowing a direct comparison of the two time-dependent approaches. The absence of electron-electron or spin-dependent interactions allows us to treat the electrons as spinless.

Following Jauho \textit{et al.} \cite{Jauho93,Jauho94}, the external time dependence due to a bias is absorbed in the tunneling matrix elements and in the single-particle energies, which become time-dependent. This assumption preserves the temporal phase coherence in the leads and central device, producing interference effects.

The time-dependent current from the lead $\alpha$ to the central region C can be obtained from the time evolution of $N_\a=\sum_k c^\dagger_{k\a}c_{k\a}$. The current $J_\a(t)$ is conveniently given by
\begin{equation}
 J_{\a}(t) = \frac{2e}{\hbar}\; \textrm{Re} \left[ \sum_{k} V_{k\a}^{*}(t) G_{k\a,0}^{<}(t,t) \right] \,,
 \label{corrente}
\end{equation}
in terms of the dot-lead lesser Green's function $G^{<}_{k\a,0}(t,t')=i\langle d^{\dagger}_{0}(t')c_{k\a}(t) \rangle$. As standard, to calculate $G_{k\a,0}^<$ we use the method of equations-of-motion to obtain the time-ordered Green's function $G_{k\alpha,0}^{\rm t}(t,t')$ followed by a contour deformation. One writes $G_{k\a,0}^{\rm t}(t,t')=-i\langle T \{ c_{k\a}(t) d^{\dagger}_{0}(t')\} \rangle$ as \cite{HaugJauho}
\begin{equation}
 G^{\rm t}_{k\a,0}(t,t') = \int dt_1 g^{\rm t}_{k\a}(t,t_1)\; V^{*}_{k\a}(t_1) \; G^{\rm t}(t_1,t')
 \label{gmista} \,,
\end{equation}
 where $G^{\rm t}(t,t')=-i\langle T [d_0(t) d_0^\dagger(t') ] \rangle$ is the Green's function of the central region and $g^{\rm t}_{k\a}(t,t')=-i\langle T [c_{k\a}(t) c_{k\a}^\dagger(t')] \rangle$ is the ``free'' uncoupled Green's function of the leads.
 
For steady-state nonequilibrium transport, all involved quantities depend only on time differences. 
In this case, the time integral in Eq.~\eqref{gmista} is a simple convolution, and one can replace 
the integral equation in time by an algebraic equation in energy by a Fourier transform. Explicit 
time-dependent terms in the Hamiltonian break time-translational invariance, making necessary 
to evaluate the two-time Green's functions. 

In a similar way, one obtains the Dyson equation for the central region Green's function
\begin{equation}
G(\tau,\tau')=G_0(\tau,\tau')+ \iint d\tau_1 d\tau_2 G_0(\tau,\tau_1)\Sigma(\tau_1,\tau_2)G(\tau_1,\tau') \,,
\label{Dyson}
\end{equation}
where the self-energy $\Sigma(\tau_1,\tau_2)=\sum_{k\a}V_{k\a}(\tau_1)g_{k\sigma}(\tau_1,\tau_2)
V_{k\a}^*(\tau_2)$ describes the coupling to the contacts. Here we consider the wide-band limit (WBL), 
which captures the main physics of typical metallic electrodes while providing analytic results. 
The wide-band approximation is valid if the density of states of the leads is a slowly varying function 
of energy in the neighborhood of the resonance energies of the central device. Typically it amounts 
to neglecting the energy shift of the dot resonance and the energy dependence of the coupling. 
The retarded/advanced self-energy in the WBL reads \cite{HaugJauho} 
\begin{equation}
 \Sigma_\alpha^{\rm r, \rm a}(t,t')=\mp \frac{i}{2}\Gamma_\alpha \delta(t-t') \,, \label{eq:WBL}
\end{equation}
where $\Gamma=2\pi \sum_{\alpha} | V_\alpha |^2$. In this picture, the leads are metallic contacts with infinite bandwidths. For the interacting case, the approximation given by Eq.~\eqref{eq:WBL} is more severe, since it means that interactions are instantaneously screened. 

In the model under analysis the resonant tunneling device is suddenly taken out of equilibrium by a switch-on of a bias $\Delta_{\a}$ on the $\a$ lead. The application of an external bias produces the  formation of a dipole around the central region, which is incorporated only as a shift in the single-particle energies of the leads \cite{Jauho93,Jauho94}. There is no further time dependence stemming from the tunnel coupling, since we restricted ourselves to the wide-band limit, Eq.~\eqref{eq:WBL}. The time lapse between the lead and dot will appear in the time evolution of the observables as coherent oscillations in the short-time transients.

In the forthcoming sections we evaluate the current using the two nonequilibrium approaches discussed in the introduction.

\section{Keldysh contour}
\label{sec:Keldysh}

\begin{center}
\begin{figure}[h!]
 \centering\includegraphics[width=\columnwidth]{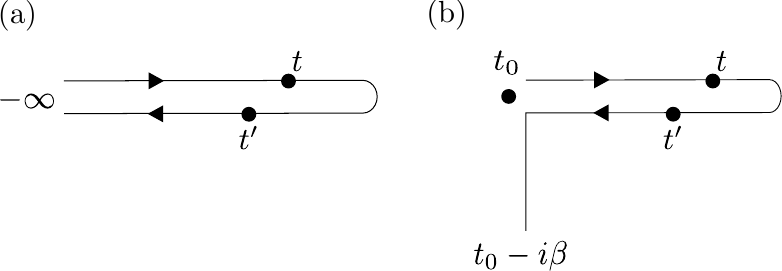}
 \caption{Contours in the complex plane, (a) Schwinger-Keldysh contour (b) extended Keldysh (Kadanoff-Baym).} 
 \label{Fig:Contour}
\end{figure}
\end{center}

The problem of calculating the objects defined in the previous section can be solved via Keldysh NEGF. In his seminal paper \cite{Keldysh}, Keldysh proposed a generalization of the diagrammatics for systems driven out of equilibrium. By defining time-ordered contour operators in the interaction representation, the expectation values of the Green's functions can be evaluated over the noninteracting states. One assumes that system starts as noninteracting in the remote past at $t_0=-\infty$ and that the interaction is slowly switched on via adiabatic hypothesis. When the system is fully interacting, the external time-dependent perturbation is applied. To avoid inconvenient integrals and, most importantly, to avoid referring to the asymptotic nonequilibrium state at $t=\infty$, the contour is folded backwards in order to switch off both perturbations returning to the noninteracting state. This procedure is equivalent to defining a two-branch time-ordered contour, exploited by Schwinger and by Keldysh \cite{Keldysh,Maciejko,SpickaRev}, illustrated in Fig.~\ref{Fig:Contour}(a).

To evaluate the time-dependent current, we need to transform integrals of two-time Green's functions 
in the complex contour of Eq.~\eqref{gmista} into an integration in the real time domain. We proceed 
according to Langreth's prescription \cite{LeeuwenIntro}, and rewrite the current in terms of the lesser, advanced and retarded contributions
\begin{equation}
 J_{\alpha}(t)=-\frac{2e}{\hbar} \mathrm{Re} \Bigl[ \Sigma_\alpha^< \cdot G^{\rm a} + \Sigma_\alpha^{\rm r} \cdot G^< \Bigr] (t,t) \,,
 \label{Kcurrent}
\end{equation}
which is commonly known as Meir-Wingreen formula for the time-independent case. In Eq.~\eqref{Kcurrent},
we adopted the short-hand notation for integrals along the Keldysh contour $ \bigl[ f \,\cdot\, g \bigr]=\int\limits_{-\infty}^\infty d\bar{t}\; f(\bar{t})\;g(\bar{t})$. 

The embedding self-energies, which incorporate the renormalization of the dot due to the 
coupling with the leads, are given by
\begin{equation}
 \Sigma^{<,\rm r}_\a(t,t')= \sum_{k} V_{k\a} g^{<,\rm r}_{k\a}(t,t') V_{k\a}^* \,, \label{eq:sigma}
\end{equation}
where we have already neglected any time-dependence in the couplings. In Eq.~\eqref{eq:sigma}, the Green's functions of the leads have a simple analytic form \cite{Jauho94}
\begin{eqnarray}
g^{<}_{k\a}(t,t')&=&i f(\varepsilon_{k\a})e^{-i\int_{t'}^{t} \varepsilon_{k\a}(t_1) dt_1}\\
g^{\rm r}_{k\a}(t,t')&=&-i \theta(t-t') e^{-i \int_{t'}^{t} \varepsilon_{k\a}(t_1) dt_1} \,.
\end{eqnarray}

As mentioned previously, in our model the application of the time-dependent bias results in a shift of the single-particle energies $\varepsilon_{k\a}(t)=\varepsilon_{k\a}+\Delta_\a(t)$. Writing the self-energies in the wide-band approximation, one obtains
\begin{eqnarray}
 \Sigma_\alpha^<(t,t')&=& i \Gamma_\alpha \int \frac{d\varepsilon}{2\pi} f_{\alpha}(\varepsilon) e^{-i\left[\varepsilon(t-t')+\int_{t'}^{t} \D_\alpha(t_1) dt_1\right]\;} \; \\
 \Sigma_\alpha^{\rm r}(t,t')&=& -i \frac{\Gamma_\alpha}{2} e^{-i \left[\varepsilon(t-t')+\int_{t'}^{t} \D_\alpha(t_1) dt_1 \right]} \delta(t-t') \,,
\label{selfenergyWBL}
\end{eqnarray}
where $\Gamma=\sum\limits_\alpha \Gamma_\alpha$ and $\Sigma^<=\sum\limits_\alpha \Sigma^<_\alpha$, and $\alpha={\textrm{L,R}}$.

The central region has the following retarded and advanced Green's functions
\begin{eqnarray}
G^{\rm r}(t,t')&=- i\theta(t-t')& e^{-i \left(\varepsilon_0 - i\Gamma/2 \right)(t-t')}  \label{GrKeldysh}\\
G^{\rm a}(t,t')&=\;\, i\theta(t'-t)& e^{-i \left(\varepsilon_0 + i\Gamma/2 \right)(t-t')} \,, \label{GaKeldysh}
\end{eqnarray}
simplified by the WBL self-energy, Eq.~\eqref{eq:WBL}.

The lesser Green's function of the dot is obtained in the integral form via Dyson's equation, Eq.~\eqref{Dyson}. By iteration and applying Langreth's rules, the Dyson's equation is rewritten as \cite{Jauho94}
\begin{align}
 G^{<}(t,t')= \, G_{in}^<(t,t') + \left[ G^{\rm r} \cdot \Sigma^{<} \cdot G^{\rm a} \right] \,, \label{GKeldysh}
\end{align}
where 
\begin{equation}
 G_{in}^<(t,t')=[ 1 + G^{\rm r} \cdot \Sigma^{\rm r}] \cdot G_0^< \cdot [ 1 + \Sigma^{\rm a} \cdot G^{\rm a} ] \label{GKeldyshin} \,.
\end{equation}

In Eq.~\eqref{GKeldysh}, the first term $G_{in}^<(t,t')$ refers to the initial free distribution, the disconnected dot. 
A possible lack of uniqueness, due to such dependence on the initial condition, was discarded by Keldysh \cite{Keldysh} 
based on the analysis of the equation of motion of $G_0^<$. 
In addition, it is expected that a heat bath washes out any dependence on the initial conditions in the remote past, 
\emph{i.e.,} for time differences much larger than the relaxation scale. 
In contrast, Caroli \textit{et al.} \cite{Caroli} considered a finite value for  $G_{in}^<(t,t')$,  attributing this feature to the lack of relaxation in their model of the leads. In our 
problem, we have explicitly verified in the time domain representation that the contractions in \eqref{GKeldyshin} 
make $G_{in}^<(t,t')$ strictly zero, helped by the singularity of the wide band approximation \eqref{eq:WBL}.
We stress that, since Keldysh's prescription assumes Dyson's equation and a well-established solution,
such as a stationary state or thermodynamical equilibrium, before the time perturbation sets in.
Thus, one should be cautious when dealing with Green's functions that violate the above conditions 
and for more general external fields, for instance, with no time translational invariance.

For a vanishing initial condition term, the correlator $G^<(t,t')$ reduces to the commonly known ``Keldysh'' 
lesser Green's function:
\begin{equation}
 G^{<}(t,t')=\iint  G^{\rm r}(t,t_1) \Sigma^<(t_1,t_2) G^{\rm a}(t_2,t') \;dt_1\,dt_2 \,,
 \label{TDGKeldysh}
\end{equation}
expected to provide the long-time transport contribution.

Let us now consider the specific case of a sudden switch-on of the bias on the $\a$ lead at $t_0=0$, namely,
\begin{eqnarray}
 \D_{\a}(t)&=& 0 \;, \quad -\infty<t<0 \nonumber \\
 &=& \D_{\a}\;, \; t\ge0 \nonumber .
\end{eqnarray}
After the perturbation, one expects to observe coherent oscillations in the $\alpha$ current $J_{\alpha}$ inversely proportional to $\D_\a$, smoothened by the coupling from the leads.

Having specified the perturbation, we can evaluate the current in Eq.~\eqref{Kcurrent}. The two corresponding convolutions result in
\begin{eqnarray}
 \Bigl[ \Sigma_\a^< \cdot G^{\rm a} \Bigr] &=& i \int \frac{d\e}{2\pi} f_\a(\e) \G_\a \Biggl[ \frac{e^{-i(\e-\e_0-i\G/2+\D_\a)t}}{(\e - \e_0 -i\G/2)} + \nonumber\\ &+& \frac{\left( 1 - e^{-i(\e-\e_0-i\G/2+\D_\a)t} \right)}{\left(\e - \e_0 -i\G/2 +\D_\a \right)} \Biggr] ,\; \label{SigmalesserGA}
 \end{eqnarray}
 and
 \begin{eqnarray}
 \Bigl[ \Sigma_\a^{\rm r} \cdot G^< \Bigr] &=& \int \frac{d\e}{2\pi} \sum_{\a'} f_{\a'}(\e) \G_\a \G_{\a'} e^{-\G t} \Biggl| \frac{1}{(\e-\e_0+i \G/2)}  \nonumber \\ 
 &+& \frac{\Bigl( e^{-i(\e-\e_0+i\G/2+\D_{\a'})t}-1 \Bigr)}{(\e-\e_0+i \G/2+\D_{\a'})} \Biggr|^2   .\; \label{SigmaAGlesser}
\end{eqnarray}

These objects have a nice interpretation: the first contribution, Eq.~\eqref{SigmalesserGA}, is related to the current flowing into the central region, while the second one, Eq.~\eqref{SigmaAGlesser}, gives the current flow from the central region to the contact $\a$. 

It is important to notice that while deriving Eq.~\eqref{SigmalesserGA} and \eqref{SigmaAGlesser} 
we performed the {unperturbed} time integrals from $-\infty$ to $0$ with the Green's function in 
\eqref{GaKeldysh}, i.e., a \textit{connected} dot. This means that, \textit{for all negative times, leads and dot are 
coupled via wide-band approximation}. To our knowledge, this step was not discussed in the literature, since the adiabatic turn-on of the couplings is assumed. Therefore we can conclude that the initial state is in equilibrium, with equal chemical potentials, and already coupled via WBL, which would dismiss the need of an adiabatic switch on of the connection. This is probably due to the fact that all electrons were considered as noninteracting and the coupling simplified to the wide-band limit, which make the problem soluble. The equivalence of the initially build-up or adiabatic coupling for the noninteracting case with relaxation is indicated by Ref.~\cite{velickyPhysicaE} by other methods.  Another delicate point is that in Ref.~\cite{Jauho94} JWM claim that the time-dependent perturbation shift is performed before the adiabatic coupling. In our interpretation, immediately before $t_0$, the system is already wide-band-coupled (interacting), with equal chemical potentials, and at $t_0$ the perturbation starts. This is equivalent to the partition-free idea.

Direct substitution of \eqref{SigmalesserGA} and \eqref{SigmaAGlesser} in \eqref{Kcurrent} results in a closed formula for the time-dependent current:
\begin{equation}
 J_\a(t)=J_\a^S +J_\a^T(t) \,,
\end{equation}
where the $J_\a^S$ is independent of time, given by
\begin{eqnarray}
 J^S_\a=-\frac{e}{\hbar} \int \frac{d{\e}}{2\pi} \G_\a \G_{\tilde{\a}} \frac{f_\a(\e-\D_\a)-f_{\tilde{\a}}(\e-\D_{\tilde{\a}})}{(\e-\e_0)^2+\G^2/4} \,,
\end{eqnarray}
since $J_\a^S=J_\a(t\gg1/\G)$ we call it the stationary current.

\begin{center}
\begin{figure}[h!]
 \centering\includegraphics[width=\columnwidth]{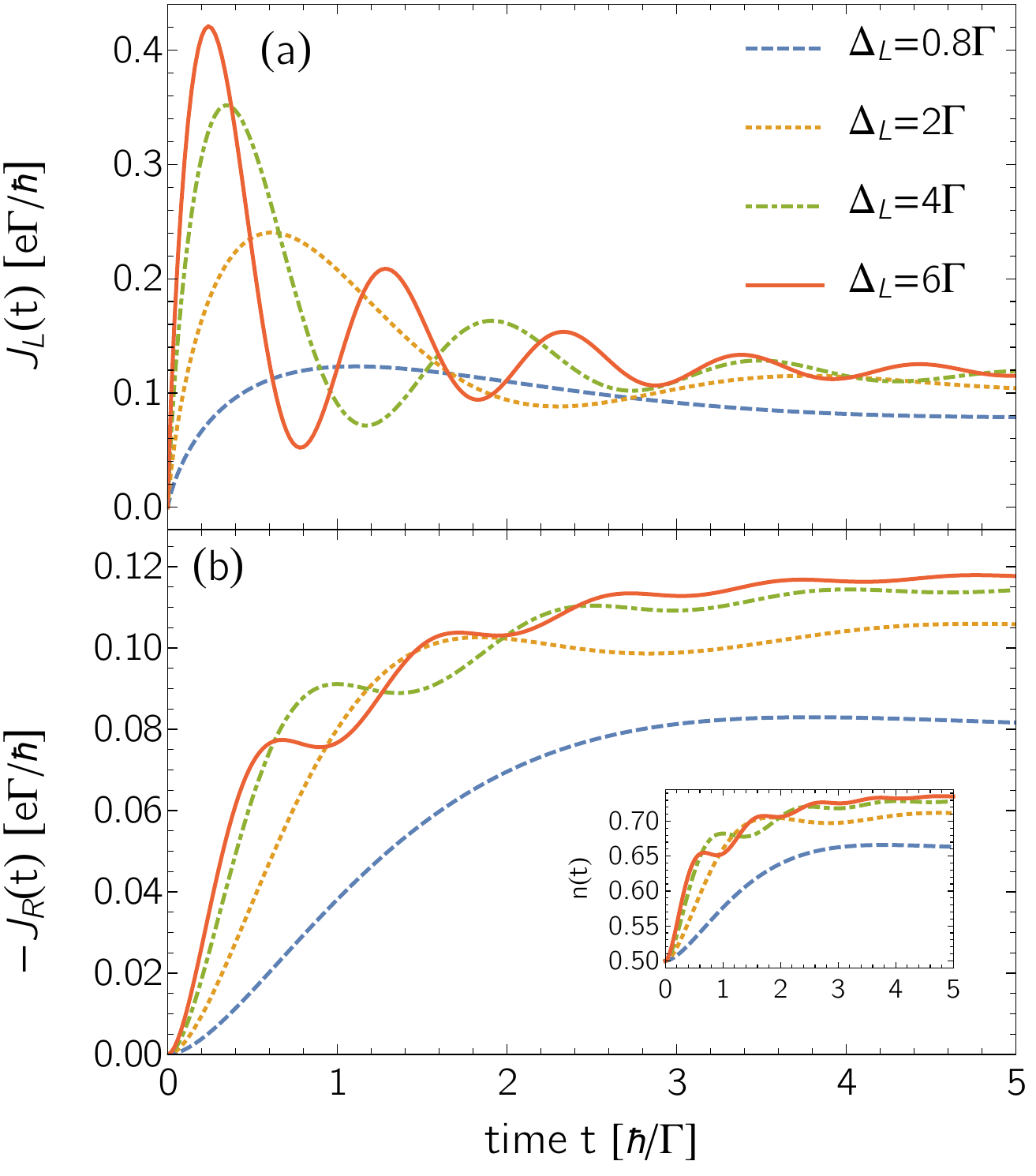}
 \caption{Time-dependent (a) left current $J_L(t)$ and (b) minus the right current $J_R(t)$ through the double-barrier tunneling device after the switch of a sudden bias $\Delta_L$ in the left lead. Note that (a) shows the same ``ringing'' behavior of Ref.~\cite{Jauho93}  (compare for instance, $\Delta_L=6\Gamma$). The inset in (b) shows the time evolution of the occupation of the dot. Numerical integration was performed at zero temperature, chemical potentials $\mu_L=\mu_R=\varepsilon_0=0$, and symmetric coupling $\Gamma_L=\Gamma_R=0.5$ ($\Gamma=\Gamma_L+\Gamma_R$).}  
 \label{Fig:Current}
\end{figure}
\end{center}

We associate the time-dependent part of $J_\a(t)$ with the transient current, that reads
\begin{eqnarray}
 &&J_\a^T(t)= \frac{2e}{\hbar} \G_\a e^{-\G t/2} \int \frac{d{\e}}{2\pi} f_\a(\e) \D_\a \Biggl\{ \Biggr. - \sum_{\a'} \frac{f_{\a'}(\e) \G_{\a'} }{2} \cdot  \nonumber \\
  &&\cdot \Biggl[ \Biggr.\frac{\D_{\a'}^2e^{-\G t/2}}{\left[(\e-\e_0)^2+\Gamma^2/4\right]\left[(\e-\e_0+\D_{\a'})^2+\Gamma^2/4\right]} + \nonumber \\
  &&+ \frac{ \D_{\a'} \Bigl[2 (\e-\e_0) {\rm cos}[(\e-\e_0+\D_{\a'})t] + \Gamma {\rm sin}[(\e-\e_0+\D_{\a'})t]\Bigr]}{{\left[(\e-\e_0)^2+\Gamma^2/4\right]\left[(\e-\e_0+\D_{\a'})^2+\Gamma^2/4\right]}}\Biggl. \Biggr] \nonumber \\
  &&- \textrm{Im}\Biggl[ \frac{e^{-i(\e-\e_0+\D_\a)t}}{(\e-\e_0-i\G/2)(\e-\e_0-i\G/2+\D_\a)} \Biggr]  \Biggl.  \Biggr\} ,
\end{eqnarray}
which 
{reproduces the expression obtained by the partition-free method \cite{Stefanucci04}. Figure \ref{Fig:Current} shows the numerical time evolution of the left, right current and dot occupation for different values of the switched bias $\Delta_L$. The ``ringing'' response of the left current, reported in Ref.\cite{Jauho93}, is also observed here, due to the phase difference between the left bias and the dot level. The larger the energy difference, the shorter is the period of oscillation. On the other hand, the right current of Ref.\cite{Stefanucci04} is also reproduced, here for values of $\Gamma_L=\Gamma_R=0.5$. Thus we have found that for this problem the Keldysh approach of Jauho, Wingreen and Meir describes transients exactly.}

To the best of our knowledge, this correspondence has not been discussed in the literature. 
The Keldysh approach was expected to reproduce the partition-free results only for very large 
time differences, \textit{i.e.,} the steady state. To achieve this full agreement in the transients, some 
approximations were crucial. 
The most evident is the absence of electron-electron interactions, which can be tackled only by a 
proper perturbation theory along the Kadanoff-Baym contour. 
Another important point is that before applying the time-perturbation, one must have an initial state 
at equilibrium: that both chemical potentials must be aligned with the level of the dot. This is the 
partition-free starting point, namely, the device is at chemical and thermal equilibrium. 
Lastly, the Markovian (or ``memory-free'') character of the wide-band coupling simplifies integrations 
from $-\infty$ to $0$ leading to the coupled initial state at $t=0$. The dot is dressed by the leads from 
$-\infty$ to $0$ by a mean-field, with trivial time/energy scales. In the next section we examine the 
partition-free approach via Kadanoff-Baym contour to have an explicit comparison of how the two 
methods develop in different contours.

\section{Extended Keldysh (Kadanoff-Baym) contour}

Another method of dealing with the expectation values in the Green's functions is defining the latter in a grand-canonical ensemble average \cite{KonstantinovPerel,KadanoffBaym}. In this description, it is standard to use Green's functions defined along the imaginary axis with periodic boundary conditions namely, Matsubara Green's functions. The advantage of such procedure is to open the possibility of considering more general initial conditions, in contrast to the Keldysh approach which assumes an uncorrelated initial condition in the remote past, as well as their influence in the very short-time transients. The mixed contour including an imaginary extension, depicted in Fig.\ref{Fig:Contour}(b), was shown to accommodate the many-body perturbation theory without need of an adiabatic hypothesis \cite{Danielewicz,Wagner}. This extended Keldysh contour is often referred in the literature as Konstantinov-Perel\cite{KonstantinovPerel}, Danielewicz \cite{Danielewicz,VelickyPRB}, and Kadanoff-Baym \cite{KadanoffBaym}. We will adopt the latter nomenclature.

The three-branch contour of Fig.\ref{Fig:Contour}(b) favors the introduction of new ``mixed'' Green's functions  with time arguments in the real and imaginary tracks. We follow the notation of Ref.~\cite{LeeuwenIntro}, namely,
\begin{eqnarray}
G^{\lceil}(\tau,t)&=&-i G^M(\tau,0)\,G^{\rm a}(t_0,t)\\
G^{\rceil}(\tau,t)&=&i G^{\rm r}(t,t_0)\,G^M(0,\tau) 
\end{eqnarray}
where $G^M$ are the Matsubara Green's functions, 
\begin{align}
G^M(\tau_1,\tau_2)=\dfrac{1}{-i\beta}\displaystyle\sum_{m=-\infty}^{\infty}\frac{{\rm e}^{-\omega_m(\tau_1-\tau_2)}}{(\omega_m-h-\Sigma^M+\mu)} \,. 
\end{align}
Integrals along the real and imaginary axis are denoted as 
$\bigl\llbracket f \cdot g \bigr\rrbracket=\int\limits_{t_0}^\infty d\bar{t}\; f(\bar{t})\;g(\bar{t})$ 
and 
$ \bigl\llbracket f \star g \bigr\rrbracket=-i\int\limits_{0}^{\beta} d\bar{t}\; f(\bar{t})\;g(\bar{t})$, 
where in the former the lower integration limit is $t_0$, this is why we change slightly the brackets 
notation to avoid confusion with the Keldysh contour. Langreth's rules are also modified, e.g., the product $c=a \cdot b$ in the Kadanoff-Baym contour has the lesser component $c^<=  a^< \cdot b^{\rm a}  + a^{\rm r} \cdot b^<  + a^\rceil \star b^\lceil $. With this in hand, we can write the current through the central device
\begin{equation}
 J_{\alpha}(t)=\frac{2e}{\hbar} \mathrm{Re} \left\llbracket \Sigma_\alpha^< \cdot G^{\rm a} + \Sigma_\alpha^{\rm r} \cdot G^< + \Sigma_\alpha^\rceil \star G^\lceil \right\rrbracket (t,t) \,,
 \label{KBcurrent}
\end{equation}
which is similar to the Keldysh current, Eq.~\eqref{Kcurrent}, except for the different integration limits and the extra term on the r.h.s.. The latter is expected to account for possible initial correlations and initial-state dependence \cite{myohanen2009kadanoff}. Note that in the noninteracting problem there are no initial correlations. The mixed self-energy contains a sum over Matsubara frequencies of the lead $\omega_q$, which in the wide-band limit results in \cite{Stefanucci-book}
\begin{equation}
 \Sigma_\alpha^\rceil(t,\tau)=\frac{\G_\a }{-i\beta} \sum_q e^{\omega_q \tau} \int \frac{ d\e}{2\pi} \frac{ e^{-i(\e+\Delta_\a)t}}{\omega_q-\e+\mu} \,.
\end{equation}

The current in Eq.~\eqref{KBcurrent} is often presented as a generalization of the Meir-Wingreen current \cite{MeirWingreen} to the transient time domain, due to the contribution $ \llbracket \Sigma_\alpha^\rceil \star G^\lceil \rrbracket$. However, we have just found in the noninteracting case, that currents from the Keldysh contour provide the same transients and steady-state from the partition-free approach of Stefanucci and collaborators\cite{Stefanucci04}, which is equivalent to integrating along the Kadanoff-Baym contour. Next we show that this contradiction is only apparent by examining how each contraction in \eqref{KBcurrent} contributes to the current. 

 For the extended contour of Fig.\ref{Fig:Contour}(b), the integrals in Eq.~\eqref{KBcurrent} become
\begin{eqnarray}
&\Bigl\llbracket \Sigma_\alpha^< \cdot G^{\rm a} \Bigr\rrbracket &= i \int \frac{d{\e}}{2\pi} f_\a(\e-\mu_\a) \G_\a  \frac{1-{\rm e}^{-i (\e-\e_0-i\Gamma/2+\D_{\a})t}}{(\e-\e_0-i\Gamma/2+\D_{\a})} \nonumber \,, \\ \label{KBSigmalesserGA}  \\
&\Bigl\llbracket \Sigma_\alpha^\rceil \star G^\lceil \Bigr\rrbracket &= i \int \frac{d{\e}}{2\pi} f_\a(\e-\mu_\a) \G_\a  \frac{{\rm e}^{-i (\e-\e_0-i\Gamma/2+\D_{\a})t}}{(\e-\e_0-i\Gamma/2)} \label{KBSigmaRightGleft} \,, \\
&\Bigl\llbracket \Sigma_\alpha^{\rm r} \cdot G^< \Bigr\rrbracket &= \frac{-i\G_\a}{2} G^<(t,t) \,. \label{KBSigmaRGlesserG}
 \end{eqnarray}
 
 First, we observe that the integration $\bigl\llbracket \Sigma_\alpha^< \cdot G^{\rm a} \bigr\rrbracket$ is not equal to $\bigl[\Sigma_\alpha^< \cdot G^{\rm a} \bigr]$ found in \eqref{SigmalesserGA}, but rather, it is the sum of Eq.~\eqref{KBSigmalesserGA} and \eqref{KBSigmaRightGleft} that reproduces Eq.~\eqref{SigmalesserGA}, the current that enters the dot. This illustrates that a direct comparison of the formulas {integrated along different contours} should be avoided.  
 Another example is the case of the ``Keldysh lesser'' Green's functions $G^{\rm r} \cdot \Sigma^< \cdot G^{\rm a}$, which will be examined later below.
 
 For the second convolution, $\bigl\llbracket \Sigma_\alpha^\rceil \star G^\lceil \bigr\rrbracket$, 
 the Matsubara sums were converted into integration along a deformed contour, indicated in 
 Refs.~\cite{Stefanucci04,Stefanucci-book}. Although the contraction runs over imaginary times,
  it yields a function of real times and pure transients,{\it  i.e.}, 
 $\lim
 _{t\to\infty}\bigl\llbracket \Sigma_\alpha^\rceil \star G^\lceil \bigr\rrbracket = 0$. 
 {Uncorrelated transients (\emph{i.e.}, produced by a noninteracting Hamiltonian), are present in all contributing terms 
 of Eq.~\eqref{KBcurrent}, not only in $\bigl\llbracket \Sigma_\alpha^\rceil \star G^\lceil \bigr\rrbracket$. (This has been also noticed in Ref.\cite{Ridley}.)}
 
 The current that leaves the dot is linked to Eq.~\eqref{KBSigmaRGlesserG}. Along the extended contour, the lesser Green's function has a more complex structure than those of the previous section. The several mixed contractions were also examined by Velicky \textit{et al.} in the study of initial correlations \cite{velickyPhysicaE} and in references therein. We will keep with the notation of Stefanucci and Almbladh \cite{Stefanucci04}. The application of Langreth's rules to the Dyson equation along the Kadanoff-Baym contour, and substitution of additional Dyson's equations, results in \cite{Stefanucci04,Stefanucci-book}
\begin{eqnarray}
 G^{<}(t,t)&=& G^{\rm r}(t,t_0) G^{<}(t_0,t_0) G^{\rm a}(t_0,t) \nonumber \\
&+&\;i\;G^{\rm r}(t,t_0)\Bigl\llbracket G^M \star \Sigma^{\lceil}\cdot  G^{\rm a} \Bigr\rrbracket(t_0,t) \nonumber \\
&-&\;i\;\Bigl\llbracket G^{\rm r}\cdot \Sigma^{\rceil}\star  G^{\rm M} \Bigr\rrbracket(t,t_0) G^{\rm a}(t_0,t) \nonumber \\
&+&\Bigl\llbracket G^{\rm r}\cdot \Sigma^{<}\cdot G^{\rm a}\Bigr\rrbracket(t,t) \nonumber \\
&+&\Bigl\llbracket G^{\rm r}\cdot\bigl\llbracket \Sigma^{\rceil}\star  G^{\rm M}\star  \Sigma^{\lceil} \bigr\rrbracket\cdot  G^{\rm a}\Bigr\rrbracket(t,t), 
\label{GlesserKB}
\end{eqnarray}
which is more intrincate than Keldysh's integral form of the Dyson's equation, Eq.~\eqref{GKeldysh}.

In Eq.~\eqref{GlesserKB}, the first term is related to the initial distribution, $G^<(t_0,t_0)=G^M(t_0,t_0^+)=\int \frac{d\zeta}{2\pi} f(\zeta)\frac{1}{\zeta-h_0}$, given by the thermodynamical ensemble.
The first and the fourth term, $G^{\rm r} \cdot \Sigma^< \cdot G^{\rm a}$, have no information about  initial correlations, indicated by the absence of the ``hooks" $\rceil,\lceil$. The second and third convolutions in Eq.~\eqref{GlesserKB} depend on the initial occupation of the dot via integrals along the imaginary track and mixed embedding $\Sigma^{\lceil,\rceil}$. The double integral in the last term of Eq.~\eqref{GlesserKB} vanishes, since the non-zero contributions from the two integrals are located in different half-planes \cite{Stefanucci04,Stefanucci-book}. 

The explicit form of each contribution of Eq.~\eqref{GlesserKB} is given, in order, by \cite{Stefanucci04} 
\begin{eqnarray}
 &&G^{<}(t,t)= i \int \frac{d{\e}}{2\pi} {\rm e}^{-\Gamma t} \Biggl\{ \frac{\Gamma \; f(\e)  }{((\e-\e_0)^2 + \Gamma^2/4)}  \nonumber \Biggr.\\
 &&+\sum_{\a} f_{\a}(\e) \G_{\a} \Biggl[ \frac{{\rm e}^{i(\e-\e_0-i\Gamma/2+\D_{\a})t}-1}{(\e-\e_0-i\Gamma/2+\D_{\a})(\e-\e_0+i\Gamma/2)}   \Biggr. \nonumber \\
 &&\qquad\qquad\qquad+\frac{{\rm e}^{-i(\e-\e_0+i\Gamma/2+\D_{\a})t}-1}{(\e-\e_0+i\Gamma/2+\D_{\a})(\e-\e_0-i\Gamma/2)} \nonumber \\
 &&+\frac{\Bigl[{\rm e}^{i (\e-\e_0-i\Gamma/2+\D_{\a})t} - 1\Bigr] \Bigl[{\rm e}^{-i (\e-\e_0+i\Gamma/2+\D_{\a})t} - 1 \Bigr]}{(\e-\e_0-i\Gamma/2+\D_{\a})(\e-\e_0+i\Gamma/2+\D_{\a})} \Biggl.\Biggr]  \Biggl.\Biggr\}\,. \nonumber\\ 
 \label{GlesserKBWBL}
\end{eqnarray}
This result is also reproduced with the Keldysh contour, Eq.~\eqref{GKeldysh}, as long both leads and dot are in thermal equilibrium, with equal chemical potentials. In Eq.~\eqref{GlesserKBWBL}, 
we can identify the unperturbed but connected dot in the denominator of the first three terms, which refer to the initial state at $t_0=0$. Although developed along different contours, the substitution of $G^<$ of Eq.~\eqref{GlesserKBWBL} back into $\bigl\llbracket \Sigma_\alpha^{\rm r} \cdot G^< \bigr\rrbracket$, in Eq.~\eqref{KBSigmaRGlesserG}, reproduces the Keldysh current leaving the dot, Eq.~\eqref{SigmaAGlesser}. 

In the long time limit the factor $e^{-\G t}$ quenches Eq.~\eqref{GlesserKBWBL}, except for the Keldysh-like convolution $\bigl\llbracket G^{\rm r}\cdot \Sigma^{<}\cdot G^{\rm a}\bigr\rrbracket$. This {integral} contributes, together with Eq.~\eqref{KBSigmalesserGA}, to the formation of the steady state current, since
\begin{eqnarray}
  \lim\limits_{t\to\infty}\Bigl\llbracket \Sigma_\alpha^{\rm r} \cdot G^< \Bigr\rrbracket &=& \frac{\G_\a }{2\pi}\sum_{\a'}\frac{\G_{\a'}\;f_{\a'}(\e-\D_{\a'})}{\e^2+\G^2/4} \,,\\
 \lim\limits_{t\to\infty}\Bigl\llbracket \Sigma_\alpha^< \cdot G^{\rm a} \Bigr\rrbracket &=& - \frac{\G}{2\pi}\frac{\G_\a f_\a(\e-\D_\a)}{(\e^2+\G^2/4)} \,.
\end{eqnarray}

In the presence of relaxation, properties of the initial state are expected to be washed out at late times of the process, which is verified by the quenching of $G^<$. In this limit {both currents converge to the steady-state.}

\section{Conclusions}
\label{sec:conclusions}

In this article we have investigated two nonequilibrium Green's functions approaches to the problem 
of a central quantum dot connected to two metallic leads. Our aim is use this simple model to compare 
the electronic transport results using the state-of-the art Keldysh approach of Jauho, Wingreen, and 
Meir \cite{Jauho94}, with those obtained from the so-called partition-free extended Keldysh approach 
of Stefanucci and Almbladh \cite{Stefanucci04}. For the case of interacting electrons, initial-state 
correlations lead to differences in the time transients \cite{myohanen2009kadanoff, Dare, Schmidt}. 
In contrast, within the single-particle approximation we find that, contrary to previous claims in the 
literature, the two approaches lead to identical results in the wide-band approximation. 
This limit leads to a closed solution of the Dyson's equation. Despite the presence of additional contractions 
along the imaginary axis of the extended Keldysh contour \textit{e.g.}, in Eq.~\eqref{GlesserKB}, they unfold 
to the same Keldysh expressions. This raises a flag of caution regarding straightforward comparison of 
formulas evaluated along different contours. We believe that the corrections in more realistic models, 
beyond the wide-band limit, are small, as long as the energy dependency in the density-of-states of the 
leads does not introduce an additional energy scale of the order of the transient time. {Attempts 
to insert an energy dependence in the couplings can be found in Refs.~\cite{Schmidt, Maciejko06}.}

We also call attention to the fact that the JWM Keldysh result rely on a WBL-\textit{connected} dot for all times before the perturbation is turned on, which contrast with the view of an adiabatic turn-on of the couplings between isolated partitions. We consider that for this particular problem of noninteracting electrons with a wide-band coupling to the leads, there is no need for the adiabaticity hypothesis, since the problem is analytically soluble from beginning. Another point of interest is that the system is connected and \text{in equilibrium} before the time perturbation, with equal chemical potentials in the central region and leads. This supports a ``partition-free'' reinterpretation of JWM's approach. 

A generalization of the problem, e.g., the incorporation of electronic interactions even at an approximate
 level or more realistic model for the junctions with couplings beyond-WBL, raises the question whether  
the initial condition term $G_{in}^<$ {is zero}  
for more intricate approximate self-energies that do not properly satisfy Dyson's equation or Keldysh's assumptions.
It has been shown that correlations modify the short-time transient profiles \cite{myohanen2009kadanoff, Dare, 
Schmidt}, as well as finite-bandwidths \cite{Maciejko06}. 

These issues are of central interest for the theory of time-dependent transport. We believe that our study 
sheds some light on the literature current controversies.

\acknowledgments
The authors acknowledge the financial support of the Brazilian funding agencies CNPq and FAPERJ.

\appendix
\section{Notation}

In this paper we use the following shorthand notation for the convolution time integrals:
\begin{eqnarray}
 \bigl[ f \,\cdot\, g \bigr](t,t') &=&\int\limits_{-\infty}^\infty d\bar{t}\; f(t,\bar{t})\;g(\bar{t},t')\\
 \bigl[ f \star g \bigr](t,t')&=&-i\int\limits_{0}^{\beta} d\bar{t}\; f(t,\bar{t})\;g(\bar{t},t') \,.
\end{eqnarray}

To compact the expressions Ref.~\cite{Jauho93} introduces the time-dependent spectral function $A(\varepsilon,t)$ 
\begin{equation}
 A_{\a}(\varepsilon,t)=\int\limits_{-\infty}^{t} {\rm e}^{i \varepsilon(t-t_1) + i\int_{t_1}^{t}\D_{\a}(t_2) dt_2 } G^r(t,t_1) \label{A} dt_1 \,.
\end{equation}

In the time-independent case ($\D_{\a}=0$), $A_{\a}$ is just the Fourier transform of the retarded Green's function, reducing to the standard spectral function. To reproduce the closed analytical expression\cite{Jauho93,Jauho94} for a sudden step bias,
\begin{equation}
 A(\e,t)=\frac{1}{\e-\e_0+i\G/2}\left\{ 1+ \D_\a\frac{1-e^{i\left(\e-\e_0+i\G/2+\D_\a \right)}}{\e-\e_0+i\G/2+\D_\a} \right\} \,,
\end{equation}
one needs to consider a connected dot before the perturbation. In addition, it was claimed\cite{Jauho94} that the perturbation shift was performed first, and the adiabatic coupling later. However the perturbation starts from a certain time $t_0$, when the system is already wide-band-interacting. This message might be significant for several Keldysh applications on tunnel devices.

In this notation, the current is written in the common form
\begin{eqnarray}
J_{\a}(t)&=&-\frac{e}{\hbar} \G_{\a} \left[ \int \frac{{d\varepsilon}}{\pi} f_{\a}(\varepsilon) \,\textrm{Im}[{ A_{\a}(\varepsilon,t) }] + n(t)  \right]  \label{JA} \\
n(t)&=&\sum_{\a'} \G_{\a'} \int \frac{d{\varepsilon}}{2\pi} f_{\a'}(\varepsilon) \bigl|A_{\a'}(\varepsilon,t)\bigr|^2 \,. \label{occ}
\end{eqnarray}
where the out contribution is written explicitly as a time-dependent occupation $n(t)$ of the central device.

\bibliography{ref}

\end{document}